\documentclass[
    reprint,
    superscriptaddress,
    amsmath,
    amssymb,
    aps,
    prl,
    floatfix,
    citeautoscript,
    longbibliography
]{revtex4-2}

\usepackage{graphicx}
\usepackage{dcolumn}
\usepackage{bm}
\usepackage{braket}
\usepackage{epstopdf}
\usepackage{color}
\usepackage{siunitx}
\usepackage{csquotes}
\usepackage{array}
\usepackage{siunitx}
\usepackage[colorlinks, linkcolor=blue, citecolor=blue, urlcolor=blue, breaklinks=true]{hyperref}
\usepackage[dvipsnames]{xcolor}
\usepackage[normalem]{ulem}

\graphicspath{{./figures/}}
\begin{document}

\title{Violation of Bell Inequality with Time--Energy Entangled Photons over a 10.2~km Free-Space Link}

\author{Lukas Bulla}
\email{lukas.bulla@univie.ac.at}
\affiliation{Institute for Quantum Optics and Quantum Information (IQOQI), Austrian Academy of Sciences, Boltzmanngasse 3, 1090 Vienna, Austria}
\affiliation{Vienna Center for Quantum Science and Technology (VCQ), Faculty of Physics, University of Vienna, Boltzmanngasse 5, 1090 Vienna, Austria}
\affiliation{Quantum Technology Laboratories GmbH -- qtlabs, Clemens-Holzmeister-Straße 6/6, 1100 Vienna, Austria}

\author{Robert Kindler}
\affiliation{Institute for Quantum Optics and Quantum Information (IQOQI), Austrian Academy of Sciences, Boltzmanngasse 3, 1090 Vienna, Austria}
\affiliation{Vienna Center for Quantum Science and Technology (VCQ), Faculty of Physics, University of Vienna, Boltzmanngasse 5, 1090 Vienna, Austria}

\author{Oskar Kohout}
\affiliation{Institute for Quantum Optics and Quantum Information (IQOQI), Austrian Academy of Sciences, Boltzmanngasse 3, 1090 Vienna, Austria}
\affiliation{Fraunhofer Institute for Applied Optics and Precision Engineering IOF,Albert-Einstein-Strasse 7, 07745 Jena, Germany}
\affiliation{Friedrich-Schiller-Universität Jena FSU, Fürstengraben 1, 07743 Jena, Germany}

\author{Kristian Hjorth}
\affiliation{Institute for Quantum Optics and Quantum Information (IQOQI), Austrian Academy of Sciences, Boltzmanngasse 3, 1090 Vienna, Austria}
\affiliation{ 
Department of Physics, Faculty of Natural Sciences, Norwegian University of Science and Technology (NTNU), NO-7491 Trondheim, Norway}

\author{Sebastian Ecker}
\affiliation{Institute for Quantum Optics and Quantum Information (IQOQI), Austrian Academy of Sciences, Boltzmanngasse 3, 1090 Vienna, Austria}
\affiliation{Vienna Center for Quantum Science and Technology (VCQ), Faculty of Physics, University of Vienna, Boltzmanngasse 5, 1090 Vienna, Austria}

\author{Sebastian P. Neumann}
\affiliation{Institute for Quantum Optics and Quantum Information (IQOQI), Austrian Academy of Sciences, Boltzmanngasse 3, 1090 Vienna, Austria}
\affiliation{Vienna Center for Quantum Science and Technology (VCQ), Faculty of Physics, University of Vienna, Boltzmanngasse 5, 1090 Vienna, Austria}

\author{Julius Bittermann}
\affiliation{Institute for Quantum Optics and Quantum Information (IQOQI), Austrian Academy of Sciences, Boltzmanngasse 3, 1090 Vienna, Austria}
\affiliation{Vienna Center for Quantum Science and Technology (VCQ), Faculty of Physics, University of Vienna, Boltzmanngasse 5, 1090 Vienna, Austria}

\author{Martin Bohmann}
\email{martin.bohmann@qtlabs.at}
\affiliation{Institute for Quantum Optics and Quantum Information (IQOQI), Austrian Academy of Sciences, Boltzmanngasse 3, 1090 Vienna, Austria}
\affiliation{Vienna Center for Quantum Science and Technology (VCQ), Faculty of Physics, University of Vienna, Boltzmanngasse 5, 1090 Vienna, Austria}
\affiliation{Quantum Technology Laboratories GmbH -- qtlabs, Clemens-Holzmeister-Straße 6/6, 1100 Vienna, Austria}

\author{Rupert Ursin}
\email{Rupert.Ursin@qtlabs.at}
\affiliation{Quantum Technology Laboratories GmbH -- qtlabs, Clemens-Holzmeister-Straße 6/6, 1100 Vienna, Austria}

\author{Marcus Huber}
\email{marcus.huber@tuwien.ac.at}
\affiliation{Vienna Center for Quantum Science and Technology, Atominstitut, TU Wien,  1020 Vienna, Austria}
\affiliation{Institute for Quantum Optics and Quantum Information (IQOQI), Austrian Academy of Sciences, Boltzmanngasse 3, 1090 Vienna, Austria}

\begin{abstract}
We report a statistically significant violation of the Bell--CHSH inequality using time--energy entangled photons distributed over a 10.2~km urban free-space link. Unlike previous field implementations, our setup employs active interferometric phase scanning without coincidence-frame binning, maintaining phase stability throughout the measurement. The raw interference visibility reaches 78\%, and we obtain an $S$-parameter of 
$S = 2.26 \pm 0.06$, exceeding the local-realistic bound by 4.3 standard deviations. Our results demonstrate the robustness of time--energy entanglement under atmospheric turbulence and mark a step towards more device-independent protocols over metropolitan distances, with link parameters emulating satellite-based quantum communication networks using phase--time encoding.
\end{abstract}

\maketitle

\begin{figure*}[t]
  \centering
  \includegraphics[width=\textwidth]{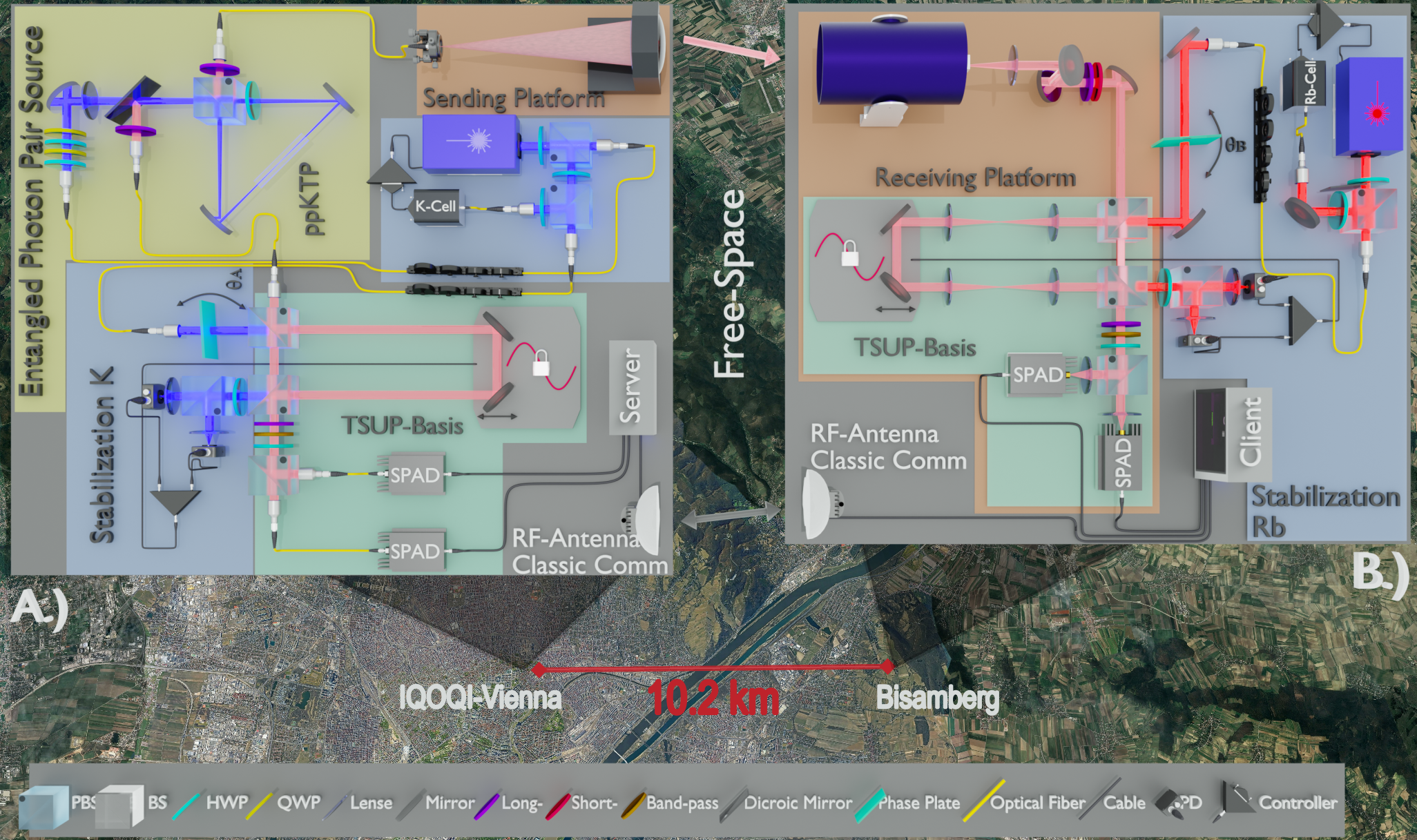}
  \caption{\textbf{Experimental configuration of the long-distance time--energy Bell test.}
  (A)\,\emph{Transmitter (Alice):} A bidirectionally pumped Sagnac source
  produces time--energy entangled photon pairs. One photon is analysed
  locally, while its partner is coupled into a fibre-fed telescope mounted
  on a computer-controlled hexapod and dispatched through an urban
  free-space link. Alice's temporal-superposition interferometer is
  phase-stabilised with a potassium reference beam.
  (B)\,\emph{Receiver (Bob):} A counterpart telescope on a hexapod collects
  the transmitted photons, which traverse a temporal-superposition
  interferometer locked to a rubidium reference before detection by
  single-photon avalanche diodes. Bob returns his time tags and basis
  settings to Alice via a radio-frequency link, where coincidences and the
  Bell parameter are evaluated.}
  \label{fig:setup}
\end{figure*}

\section{Introduction}

Bell’s theorem transformed our understanding of quantum mechanics by showing that no local-realistic model can reproduce the correlations exhibited by entangled particles. 
Decisive experimental violations of Bell inequalities by Clauser, Aspect, and Zeilinger, honored with the 2022 Nobel Prize in Physics, have cemented entanglement 
as a cornerstone of modern physics \cite{clauser72,Aspect82,zeilinger98}. 
Beyond a fundamental test of non-local realism, the emergence of quantum technologies provided a renewed motivation for an experimental study of Bell inequalities. Violating a Bell inequality is equivalent to a test of entanglement, making no assumptions on the local measurement devices or the Hilbert space dimension underlying the measurement model. 
Comprehensive reviews trace a half-century of increasingly sophisticated tests \cite{Brunner14}, the vast majority exploiting polarization-entangled photon pairs because of their experimental accessibility and high interference contrast \cite{Giustina_Belltest2015,Rowe2001,Valles2014,schuck06}. In the context of quantum key distribution (QKD), this allows for realising the fullest promise of quantum communication with security proven in a completely device independent manner\cite{acin2006bell}. While for polarisation the Hilbert space is small and measurement devices are well-understood, more promising directions overcoming the intrinsic limitations of qubit based communication are provided by continuous degrees of freedom that can be discretised into high-dimensional Hilbert spaces. This paradigm of high-dimensional quantum communication promises to not only increase key rates per entangled pair detection, but provide an intrinsic noise resistance that make it ideal candidates for long-distance and satellite based quantum communication~\cite{ecker19, ecker2021experimental}.
And in such continuous degrees of freedom, Bell inequalities provide an even more important tool for entanglement certification, as the process of discretisation opens many new exploitable assumptions that would not occur in an inherently finite dimensional setting.
 Violations of Bell inequalities have been reported in orbital-angular-momentum, path, position-momentum, and time–energy degrees of freedom \cite{rarity90,kwiat93,Santagiustina24,mair2001}. In the latter, a Franson interferometer consisting of two imbalanced Mach-Zehnder interferometers enables nonlocal two-photon interference, provided the photons are emitted within a coherence-time window much shorter than the interferometer imbalance \cite{franson89}. Such time–energy entanglement arises naturally in continuous-wave pumped spontaneous parametric down-conversion and supports high-dimensional encoding for quantum communication \cite{erhard2018,kues2017,xie2015, bulla2023genuinehighdim}. Laboratory implementations have demonstrated high-visibility Franson interference over fiber~\cite{Tittel1998,Santagiustina24} and short free-space \cite{steinlechner17} channels, yet long-distance links remain challenging owing to phase drifts, atmospheric turbulence\cite{vasylyev2012toward,klen2023numerical,Rossi25}, and accidental coincidences that erode the Bell correlation. Overcoming these obstacles is essential for deploying entanglement-based links of a sufficient quality to violate Bell inequalities.
From a quantum communication perspective this is essential to further increase the security of entanglement-based protocols, as the violation of Bell inequalities rules out almost any hardware-based attack, ie. entering the so called partially device independent regime. While a fully device-independent implementation would require closing all of the relevant loopholes, long distance links will still require coincidence postselection and thus leaving the fair sampling assumption open. No assumptions, however, are needed on the instrinsic dimensionality, the discretisation procedure or the unbiasedness of the measurement settings.

In this work, we perform a Bell--CHSH test using a Franson-type setup on a \SI{10.2}{km} free space channel, already used for proof-of-principle QKD under similar conditions~\cite{bulla2023PRX}.
We actively scan the interferometric phase, apply no post-processing beyond post-selection on coincidence counts, and quantify Bell violations directly.
To this end, we expand and build on a previous quantum-communication implementation to perform a Bell--CHSH violation over a long-distance free-space link in the time-energy DoF.
In addition to providing another datapoint useful for fundamental tests of physics, we discuss potential expansions of our work and its relevance for applications in quantum technology applications.

Our setup builds directly on the phase-stable architecture established in~\cite{bulla2023distribution}, with the addition of critical modifications tailored for a direct Bell inequality test.

The photon-pair source is driven by a continuous-wave diode laser at \SI{404.53}{nm}, locked to a \({}^{39}\mathrm{K}\) transition \cite{Hanley_2015}. It pumps a \SI{30}{mm} type-II PPKTP crystal in a bidirectional Sagnac loop, yielding spectrally degenerate photon pairs at \SI{808.9}{nm}. The long crystal yields a narrow single-photon coherence time of $\sim$\SI{3}{ps},ensuring high entanglement visibility due to intrinsic pair indistinguishability.

One photon is retained locally at Alice's station, while its partner is sent to Bob across a \SI{10.2}{km} free-space channel traversing urban airspace. Each endpoint employs hexapod-mounted telescopes (\SI{350}{mm} Newtonian at Alice, 254~mm Cassegrain at Bob) that track bidirectional \SI{532}{nm} beacons to maintain sub-\SI{5}{\micro\radian} pointing errors. This stabilization yields an average link attenuation of 23--\SI{25}{dB} under urban seeing conditions.

At each site, a free-space Mach--Zehnder interferometer with a path-length difference of \SI{80.8}{cm} (\SI{2.7}{ns}) converts the measurement into a Franson-type interferometric configuration without requiring post-selection \cite{bulla2023distribution}. To suppress wavefront distortion due to atmospheric turbulence, Bob's long arm includes two cascaded 4\emph{f} systems~\cite{jin18}  . Active phase stabilization is achieved via a \SI{780.23}{nm} laser, locked to a \({}^{87}\mathrm{Rb}\) transition \cite{preston1996doppler}, independently of the pump. The stabilization signal is extracted from interference fringes and used to drive piezo-mounted mirrors, achieving sub-micro-radian phase noise over the 1--\SI{100}{s} band.

Our implementation introduces a deterministic, actively tunable phase reference by rotating a plate in the stabilization beam path. This approach enables precise, repeatable control of the interferometer phase at the quantum wavelength (\SI{808.9}{nm}) via the lock beam (\SI{780}{nm}). Crucially, this allows us to continuously scan the nonlocal phase without interrupting data acquisition on both interferometers.

Coincidence events are recorded using free-running silicon SPADs with quantum detection efficiencies near 70\% and dark-count rates below {\SI{1400}{Hz}}. Time tags are acquired with \SI{1}{ps} resolution, and the independent clocks are synchronised to \(\sim\)\SI{20}{ps} via two GPS-disciplined rubidium clocks as well as real-time cross-correlation of the raw time streams obtained by the entangled photons correlated arrival times~\cite{bulla2023distribution}.
Departing from earlier high-dimensional demonstrations, we confine the analysis to a two-dimensional time-bin encoding, registering coincidences within a single \SI{1}{ns} window. The nonlocal phase \(\phi_A\) at Alice was scanned over a range of slightly more than $3\pi/4$ in adaptive steps (2° in the wings, 0.5° near the values required to maximally violate the CHSH-inequality). Each phase point was integrated for \SI{30}{s}, after which the lock was verified before proceeding. The full scan was then repeated with Bob's phase shifted by $\pi/2$.
This measurement strategy, together with our robust interferometric phase control and turbulence mitigation, enables reliable high-visibility interference, forming the basis of the Bell test described in the following.

\begin{figure}[t]
    \centering
    \includegraphics[width=0.5\textwidth]{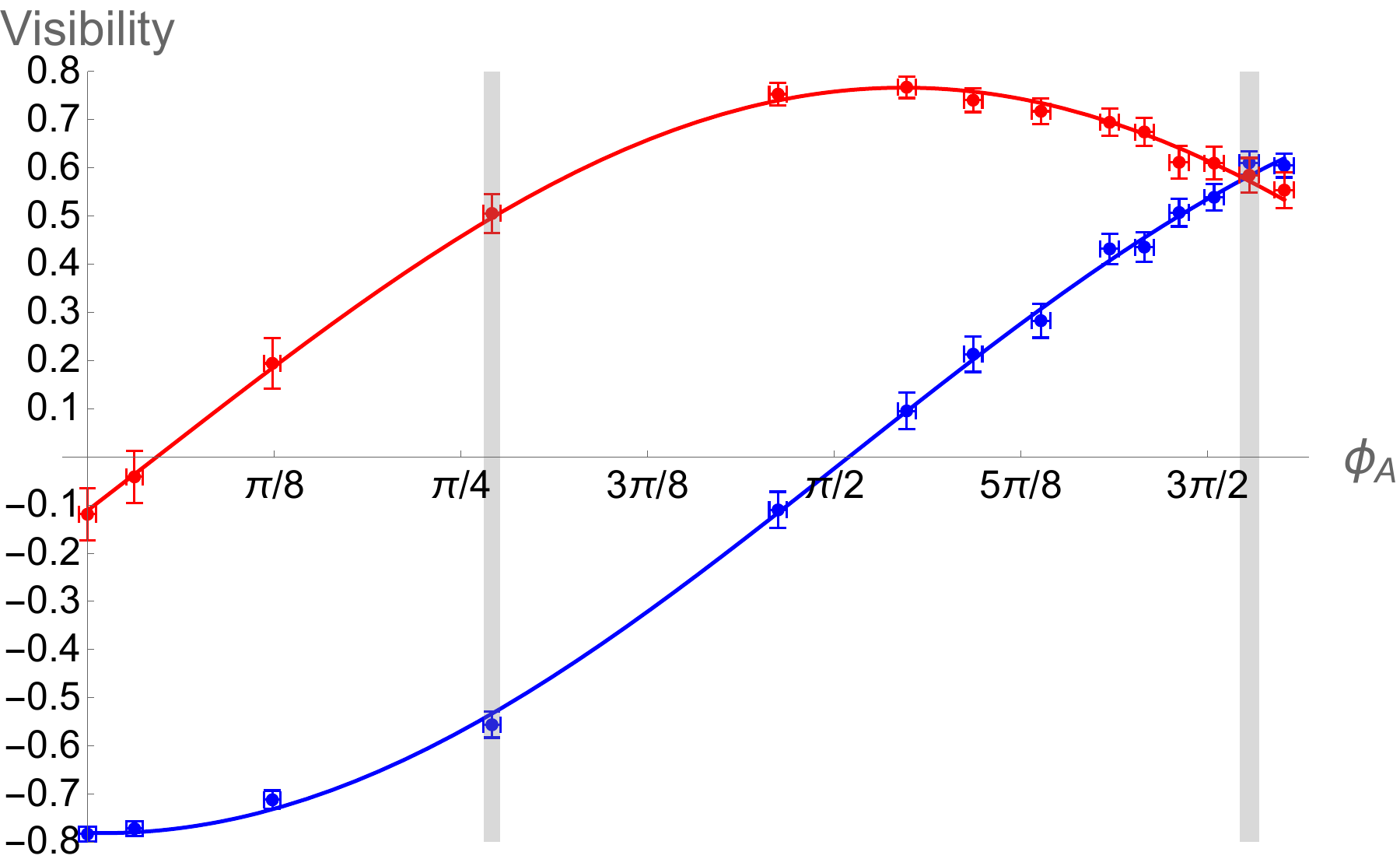}
    \caption{%
    \textbf{Two-photon interference fringes.}\,
    Displayed are the measured correlation coefficients
    $E\!\bigl(\phi_{A},\phi_{B}\bigr)$ (blue symbols) and $E\!\bigl(\phi_{A},\phi_{B}+\pi/2\bigr)$ (red symbols) as functions of Alice’s relative phase~$\phi_{A}$.  For the blue data set Bob’s phase $\phi_{B}$ is kept at its nominal value, whereas for the red data set it is shifted by~$\pi/2$.  Phase control is realized by tilting a calibrated wave-plate in the actively stabilized interferometer of each station; the measured tilt angles are converted into phase shifts on the abscissa. On top of Alice’s phase $\phi_{A}$, the x-axis also shows an offset caused by other optical components.
    Error bars correspond to $\pm1\sigma$ and are obtained by propagating Poissonian counting statistics in conjunction with shot-to-shot intensity fluctuations of the free-space channel over the measurement time.
    Solid curves are independent least-squares sinusoidal fits and serve only
    as guides to the eye.  The four shaded markers identify the phase settings used to evaluate the CHSH parameter~$S$.  The figure spans both columns to make the high-visibility fringe structure clearly discernible to the reader.}
    \label{fig:fringes}
\end{figure}

\section{Results}
We performed a CHSH-type Bell test by explicitly measuring the expectation values
\begin{equation}
E(\phi_A, \phi_B)=\bra{\psi}\hat{\sigma}_{\phi_A}\otimes \hat{\sigma}_{\phi_B}\ket{\psi}
\end{equation}
using the Pauli operator
\begin{equation}
\hat{\sigma}_{\phi_A}=\vec{n}\hat{\sigma}=\sin(\phi_A)\hat{\sigma}_x+\cos(\phi_A)\hat{\sigma}_z
\end{equation}
Experimentally, these expectation values are calculated from the coincidence-click correlations (CC) between different detectors, labeled $1,2$ for Alice and $3,4$ for Bob:
\begin{equation}
\text{Vis}_{A,B}=\frac{\text{CC}(1,3)+\text{CC}(2,4)-\text{CC}(1,4)-\text{CC}(2,3)}{\text{CC}(1,3)+\text{CC}(2,4)+\text{CC}(1,4)+\text{CC}(2,3)} 
\end{equation}
To find the best combination of phases suitable for a Bell test, we first fixed Bobs phase at its lowest possible value, setting the respective phase-plate to $0^\circ$, while scanning Alices phase over a range of more than $3\pi/4$, followed by repeating the same measurements with Bobs phase now increased by $\pi/2$. 
At each of these phase combinations, we measured the coincidence rates for 30~s and computed the corresponding correlation values $E(\phi_A,\phi_B)$ from the raw data (See Fig. \ref{fig:fringes}). We observed a residual phase $\phi$, originating from additional phase shifts from reflections inside our beam path, effectively causing an offset on the x-axis in  Fig. \ref{fig:fringes}, which does not have any negative effect on our data quality.
As expected, the two best-fitting datapoints on each curve show a phase shift of $\pi/2$ and the cosine fit reveals that the actual phase shift between both graphs is actually $0.53\pi$ instead of the anticipated value of $\pi/2$.
Four of these data points were thus identified and allowed us to evaluate the Bell-CHSH parameter in a proper black-box paradigm.

\begin{equation}
S = |E(\phi_A, \phi_B) - E(\phi_A, \phi_B') + E(\phi_A', \phi_B) + E(\phi_A', \phi_B')| .
\end{equation}
We obtained

\begin{equation}
S = 2.26 \pm 0.06
\end{equation}
exceeding the classical limit of 2 by 4.3 standard deviations, thereby significantly violating the CHSH inequality.  The raw fringe visibility across the full phase scan averaged $78\%$, indicating stable interference and consistent phase coherence of the quantum state, despite the detrimental effects of the free space link.
Importantly, the result is derived without background subtraction or post-selection. All detector clicks within the \SI{1}{ns} window contribute to the statistics.

The observed visibility  and hence the Bell parameter S is mainly limited by multi-pair emission, detector  after-pulsing  and dark counts, residual phase noise, optical imperfections, background light, and spatial misalignment. Additional contributions arise from non-ideal beam splitter ratios, scattering losses, and incomplete turbulence compensation, all of which reduce the coherence and contrast of the interference fringe.
Our analysis confirms that the observed Bell violation originates from genuine time--energy entanglement preserved across a \SI{10.2}{km} free-space link, even under strong atmospheric turbulence and background light conditions.
\section{Discussion and Outlook}

The observed violation of the Bell inequality in our time--energy entangled system over a \SI{10.2}{km} free-space link demonstrates the feasibility of deploying quantum non-locality tests and entanglement-based quantum protocols under real-world atmospheric conditions. Our results validate and reinforce the robustness of time--energy entanglement as a reliable carrier of quantum information in metropolitan-scale environments. This may also be of fundamental interest, as time itself isn't an observable in the quantum sense and the time-energy uncertainty is qualitatively different from position-momentum argued to be the basis for spatial entanglement generation.

Unlike previous demonstrations on this link~\cite{bulla2023genuinehighdim} focused on high-dimensional encoding, our experiment the binary aspect of entanglement but is able to provide a full CHSH test. In the future, it would be of great interest to combine the two avenues and explore which Bell tests would be suitable for Franson type setups and genuine high-dimensional encoding.

Further improvements in detection efficiency, e.g., through adaptive optics and deployment of superconducting nanowire single-photon detectors, could move the present configuration closer to a loophole-free regime. 
In addition, the demonstrated stability and scalability of our actively phased Franson interferometers suggest the practicality of real-time entanglement distribution for quantum networking. The deterministic phase-scan architecture may be extended to encode or track quantum states dynamically, enabling high-dimensional protocol implementations \cite{floalexhighdim}.

These advances position time--energy entanglement as a key enabling technology for next-generation quantum communication networks. Combined with satellite downlinks \cite{dequal2016experimental,yin17}or fiber-based \cite{Wengerowsky19,neumann2022continuous}   channels, they pave the way toward global-scale, high-rate, entanglement distribution infrastructure.

\section{Conclusion}

We have reported a Bell--CHSH violation using time--energy entangled photon pairs distributed over a 10.2~km turbulent free-space link, achieving a raw CHSH parameter of $S = 2.26 \pm 0.06$ without coincidence-frame binning or post-selection. This result exceeds the classical limit by 4.3 standard deviations, and since for the chosen angles $\sqrt{2}-S$ is a tight entanglement witness, it also confirms the presence of entanglement by over 14~\(\sigma\).

Our implementation leverages deterministic, actively controlled phase scanning and robust interferometric stabilisation to maintain high visibility interference in real-world atmospheric conditions. The measurement confirms the practical utility of time--energy entanglement as a resource for long-distance quantum communication and lays the groundwork for future tests of nonlocality in free space with custom partial device independent features.

Together with prior high-dimensional demonstrations, our results advance the experimental state of the art and provide a realistic performance benchmark for future quantum networking protocols, especially those seeking minimal assumptions in untrusted scenarios.


{\bf Acknowledgments}. M.H. acknowledges support from  the Horizon-Europe research and innovation programme under grant agreement No 101070168 (HyperSpace).
\bibliography{references}

\clearpage

\section{\label{sec:Methods}Appendix}

\subsection{Active Phase Control and Deterministic Phase Scanning}
\label{sub:Phase Control}

Precise, deterministic control of the nonlocal phase between Alice’s and Bob’s interferometers is critical for scanning interference fringes without disturbing the post-selection-free Franson configuration. Our scheme extends beyond passive stabilisation by directly controlling the phase via a  birefringent element acting on the stabilisation beam.

At Alice's station, the unbalanced Mach--Zehnder interferometer is actively stabilised using the residual pump light at \SI{404.53}{nm}, which is frequency-locked to the \({}^{39}\mathrm{K}\,4S_{1/2}\rightarrow5P_{3/2}\) transition. At Bob’s station, stabilisation is achieved independently with a \SI{780.23}{nm} diode laser locked to the \({}^{87}\mathrm{Rb}\,5S_{1/2}\rightarrow5P_{3/2}\) transition. In both cases, the stabilisation light enters in a diagonal polarization state via the unused input port of the first polarising beam splitter (PBS) of the Mach--Zehnder, co-propagating with the quantum signal through both interferometric arms. At the second PBS both signals are separated due to their polarization state. At this point, no Mach–Zehnder or Franson interference is visible due to the photons’ orthogonal polarization states, which leave them distinguishable. Therefore, the polarization is rotated into the diagonal basis and then sent to another PBS, making them indistinguishable. 
Due to the quantum signal's small coherence length, no Mach--Zehnder interference can be observed in that path and the effects of phase changes inside the interferometer will only become apparent when looking at two-photon coincidences. 
In contrast to that, the stablilization signal shows well known Mach-Zehnder interference. 
The phase \(\phi\) is tuned by tilting a multi-order half-wave plate placed in the stabilization path before the interferometers. The wave plate introduces a differential optical path delay between horizontal and vertical polarisations, which upon tilt by an angle \(\theta\) changes according to
\begin{equation}
\Delta\phi(\theta) = \frac{2\pi}{\lambda} \left( d_e(\theta)n_e(\theta) - d_o(\theta)n_o(\theta) \right),
\end{equation}
where \(\lambda\) is the stabilisation laser wavelength,  \(n_{e,o}(\theta)\) are the effective ordinary and extraordinary refractive indices at tilt angle \(\theta\), and \(d_{e}(\theta)\) (\(d_{o}(\theta)\)) is the distance that extraordinarily (ordinarily) polarized light travels through the plate.
As long as the plate is hit under normal incidence (i.e., $\theta=0$), both light paths are identical and equal to the waveplate's thickness.
Changing the phase will consequently alter the splitting ratio of both stabilization detectors, which are part of a PID controled feedback-lock that forces the splitting ratio to remain at $50/50$ by countering the stabilization signals phase with the piezo in the interferometers long arm. The phase introduced by the piezos will thus change the quantum signal's \"{}non-local\"{} phase and thereby the single photon detectors' correlations.


At Alice's station, the stabilisation beam at \SI{404.53}{nm} is exactly half the wavelength of the quantum signal at \SI{808.9}{nm}. Therefore, the induced optical phase shift \(\Delta\phi_{\mathrm{stab}}\) on the stabilisation beam translates linearly into a corresponding shift \(\Delta\phi_{\mathrm{quantum}}\) at the quantum wavelength:
\begin{equation}
\Delta\phi_{\mathrm{quantum}} = 2\,\Delta\phi_{\mathrm{stab}}.
\end{equation}

At Bob's station, the situation requires more careful treatment, as the stabilisation wavelength (\SI{780.23}{nm}) differs from the quantum channel wavelength (\SI{808.9}{nm}). The phase shift imparted at the stabilisation wavelength \(\Delta\phi_{\mathrm{stab}}\) translates to a quantum phase shift \(\Delta\phi_{\mathrm{quantum}}\) according to
\begin{equation}
\Delta\phi_{\mathrm{quantum}} = \Delta\phi_{\mathrm{stab}} \times \frac{\SI{780.23}{nm}}{\SI{808.9}{nm}}.
\end{equation}
This wavelength mismatch factor is calculated and accounted for during the phase scanning procedure. Tilt increments of the wave plate are chosen accordingly to achieve uniform scanning steps at the quantum wavelength.

The mapping between the stabilisation and quantum phases is verified experimentally via fringe tracking during controlled scans. At both stations, the technique enables deterministic, high-resolution phase scanning across the Bell-relevant settings

\begin{figure*}[t]
    \centering
    \includegraphics[width=\textwidth]{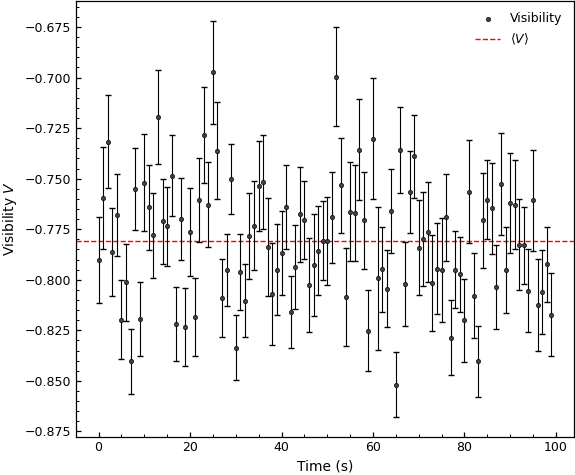}
    \caption{%
    \textbf{Time-resolved visibility measurement for a representative dataset used in the evaluation of the Bell inequality.}\,
     The figure shows the signed two-photon coincidence Visibility $(V(t))$, evaluated over 100 consecutive integration intervals of 1 s each. Black markers represent individual visibility estimates calculated from four detector coincidence channels$(C_{1},C_{2},C_{3},C_{4})$, while the vertical error bars indicate the $1 \sigma$ statistical uncertainties , by standard error propagation of independent Poisson counting uncertainties. The red dashed line denotes the arithmetic mean $(\langle V\rangle=-0.781)$. Over the 100-second acquisition interval, the measured values exhibit a sample standard deviation of $(\sigma_V=0,03)$. The figure illustrates the temporal variation and statistical properties of a single visibility measurement that contributes to the evaluation of the Bell inequality.}
    \label{fig:vissample}
\end{figure*}

\end{document}